\documentclass[twocolumn]{aastex63}

\usepackage{devanagari}
\usepackage{upgreek}
\usepackage[encapsulated]{CJK} % for chinese characters
\usepackage{ucs}
\newcommand{\cntext}[1]{\begin{CJK}{UTF8}{gbsn}#1\end{CJK}}

\received{October 23, 2020}
\revised{March 8, 2021}
\accepted{March 9, 2021}
\published{April 12, 2021}
\submitjournal{PSJ}

\shorttitle{Venus Observations}
\shortauthors{Dahal et al.}
\graphicspath{{./}{figures/}}

\begin{document}

\title{Venus Observations at 40 and 90 GHz with CLASS}

\correspondingauthor{Sumit Dahal}
\email{sumit.dahal@nasa.gov}

\author[0000-0002-1708-5464]{Sumit Dahal}
\affiliation{NASA Goddard Space Flight Center, 8800 Greenbelt Road, Greenbelt, MD 20771, USA}
\affiliation{Department of Physics and Astronomy, Johns Hopkins University, 3701 San Martin Drive, Baltimore, MD 21218, USA}

\author{Michael K. Brewer}
\affiliation{Department of Physics and Astronomy, Johns Hopkins University, 3701 San Martin Drive, Baltimore, MD 21218, USA}

\author[0000-0002-8412-630X]{John W. Appel}
\affiliation{Department of Physics and Astronomy, Johns Hopkins University, 3701 San Martin Drive, Baltimore, MD 21218, USA}

\author[0000-0001-7941-9602]{Aamir Ali}
\affiliation{Department of Physics, University of California, Berkeley, CA 94720, USA}
\affiliation{Department of Physics and Astronomy, Johns Hopkins University, 3701 San Martin Drive, Baltimore, MD 21218, USA}

\author[0000-0001-8839-7206]{Charles L. Bennett}
\affiliation{Department of Physics and Astronomy, Johns Hopkins University, 3701 San Martin Drive, Baltimore, MD 21218, USA}

\author[0000-0001-8468-9391]{Ricardo Bustos}
\affiliation{Facultad de Ingenier\'{i}a, Universidad Cat\'{o}lica de la Sant\'{i}sima Concepci\'{o}n, Alonso de Ribera 2850, Concepci\'{o}n, Chile}

\author{Manwei Chan}
\affiliation{Department of Physics and Astronomy, Johns Hopkins University, 3701 San Martin Drive, Baltimore, MD 21218, USA}

\author[0000-0003-0016-0533]{David T. Chuss}
\affiliation{Department of Physics, Villanova University, 800 Lancaster Avenue, Villanova, PA 19085, USA}

\author{Joseph Cleary}
\affiliation{Department of Physics and Astronomy, Johns Hopkins University, 3701 San Martin Drive, Baltimore, MD 21218, USA}

\author[0000-0002-0552-3754]{Jullianna D. Couto}
\affiliation{Department of Physics and Astronomy, Johns Hopkins University, 3701 San Martin Drive, Baltimore, MD 21218, USA}

\author[0000-0003-3853-8757]{Rahul Datta}
\affiliation{Department of Physics and Astronomy, Johns Hopkins University, 3701 San Martin Drive, Baltimore, MD 21218, USA}

\author{Kevin L. Denis}
\affiliation{NASA Goddard Space Flight Center, 8800 Greenbelt Road, Greenbelt, MD 20771, USA}

\author[0000-0001-6976-180X]{Joseph Eimer}
\affiliation{Department of Physics and Astronomy, Johns Hopkins University, 3701 San Martin Drive, Baltimore, MD 21218, USA}

\author{Francisco Espinoza}
\affiliation{Facultad de Ingenier\'{i}a, Universidad Cat\'{o}lica de la Sant\'{i}sima Concepci\'{o}n, Alonso de Ribera 2850, Concepci\'{o}n, Chile}

\author[0000-0002-4782-3851]{Thomas Essinger-Hileman}
\affiliation{NASA Goddard Space Flight Center, 8800 Greenbelt Road, Greenbelt, MD 20771, USA}
\affiliation{Department of Physics and Astronomy, Johns Hopkins University, 3701 San Martin Drive, Baltimore, MD 21218, USA}

\author{Dominik Gothe}
\affiliation{Department of Physics and Astronomy, Johns Hopkins University, 3701 San Martin Drive, Baltimore, MD 21218, USA}

\author[0000-0003-1248-9563]{Kathleen Harrington}
\affiliation{Department of Astronomy and Astrophysics, University of Chicago, 5640 South Ellis Avenue, Chicago, IL 60637, USA}
\affiliation{Department of Physics and Astronomy, Johns Hopkins University, 3701 San Martin Drive, Baltimore, MD 21218, USA}

\author[0000-0001-7466-0317]{Jeffrey Iuliano}
\affiliation{Department of Physics and Astronomy, Johns Hopkins University, 3701 San Martin Drive, Baltimore, MD 21218, USA}

\author{John Karakla}
\affiliation{Department of Physics and Astronomy, Johns Hopkins University, 3701 San Martin Drive, Baltimore, MD 21218, USA}

\author[0000-0003-4496-6520]{Tobias~A. Marriage}
\affiliation{Department of Physics and Astronomy, Johns Hopkins University, 3701 San Martin Drive, Baltimore, MD 21218, USA}

\author{Sasha Novack}
\affiliation{Department of Physics and Astronomy, Johns Hopkins University, 3701 San Martin Drive, Baltimore, MD 21218, USA}

\author[0000-0002-5247-2523]{Carolina N\'{u}\~{n}ez}
\affiliation{Department of Physics and Astronomy, Johns Hopkins University, 3701 San Martin Drive, Baltimore, MD 21218, USA}

\author[0000-0002-0024-2662]{Ivan L. Padilla}
\affiliation{Department of Physics and Astronomy, Johns Hopkins University, 3701 San Martin Drive, Baltimore, MD 21218, USA}

\author[0000-0002-8224-859X]{Lucas Parker}
\affiliation{Space and Remote Sensing, MS D436, Los Alamos National Laboratory, Los Alamos, NM 87544, USA}
\affiliation{Department of Physics and Astronomy, Johns Hopkins University, 3701 San Martin Drive, Baltimore, MD 21218, USA}

\author[0000-0002-4436-4215]{Matthew~A. Petroff}
\affiliation{Department of Physics and Astronomy, Johns Hopkins University, 3701 San Martin Drive, Baltimore, MD 21218, USA}

\author[0000-0001-5704-271X]{Rodrigo Reeves}
\affiliation{CePIA, Departamento de Astronom\'{i}a, Universidad de Concepci\'{o}n, Concepci\'{o}n, Chile}

\author{Gary Rhoades}
\affiliation{Department of Physics and Astronomy, Johns Hopkins University, 3701 San Martin Drive, Baltimore, MD 21218, USA}

\author[0000-0003-4189-0700]{Karwan Rostem}
\affiliation{NASA Goddard Space Flight Center, 8800 Greenbelt Road, Greenbelt, MD 20771, USA}

\author[0000-0003-3487-2811]{Deniz A. N. Valle}
\affiliation{Department of Physics and Astronomy, Johns Hopkins University, 3701 San Martin Drive, Baltimore, MD 21218, USA}

\author[0000-0002-5437-6121]{Duncan J. Watts}
\affiliation{Institute of Theoretical Astrophysics, University of Oslo, P.O. Box 1029 Blindern, N-0315 Oslo, Norway}
\affiliation{Department of Physics and Astronomy, Johns Hopkins University, 3701 San Martin Drive, Baltimore, MD 21218, USA}

\author[0000-0003-3017-3474]{Janet L. Weiland}
\affiliation{Department of Physics and Astronomy, Johns Hopkins University, 3701 San Martin Drive, Baltimore, MD 21218, USA}

\author[0000-0002-7567-4451]{Edward J. Wollack}
\affiliation{NASA Goddard Space Flight Center, 8800 Greenbelt Road, Greenbelt, MD 20771, USA}

\author[0000-0001-5112-2567]{Zhilei Xu (\cntext{徐智磊}$\!\!$)}
\affiliation{ MIT Kavli Institute, Massachusetts Institute of Technology, 77 Massachusetts Avenue, Cambridge, MA 02139, USA}
\affiliation{Department of Physics and Astronomy, Johns Hopkins University, 3701 San Martin Drive, Baltimore, MD 21218, USA}
%\nocollaboration{0}

\begin{abstract}
Using the Cosmology Large Angular Scale Surveyor, we measure the disk-averaged absolute Venus brightness temperature to be \mbox{432.3 $\pm$ 2.8}~K and \mbox{355.6 $\pm$ 1.3}~K in the \textit{Q} and \textit{W} frequency bands centered at 38.8 and 93.7 GHz, respectively. At both frequency bands, these are the most precise measurements to date. Furthermore, we observe no phase dependence of the measured temperature in either band. Our measurements are consistent with a CO$_2$-dominant atmospheric model that includes trace amounts of additional absorbers like SO$_2$ and H$_2$SO$_4$.
\end{abstract}

\keywords{\href{http://astrothesaurus.org/uat/1763}{Venus (1763)}; \href{http://astrothesaurus.org/uat/182}{Brightness temperature (182)}; \href{http://astrothesaurus.org/uat/2120}{Atmospheric composition (2120)}}

\section{Introduction} \label{sec:intro}
Microwave observations of Venus can be used to probe its hot and dense atmosphere that mostly ($\sim$~96\%) consists of CO$_2$ \citep{oyama1979}. The greenhouse effect from the thick Venusian atmosphere, which reaches $\sim$~90~bars at the surface, maintains the surface temperature at $\sim$ 750 K \citep{muhleman1979}. While radio wavelengths $\gtrsim$~4~cm probe the surface, decreasing wavelengths successively probe increasing altitudes in the atmosphere with a steep decrease in temperature \citep{dePater1990, butler2001}. The measurement of the brightness temperature and its phase dependence at different microwave frequencies can therefore reveal important information about the composition and dynamics of various layers of the Venusian atmosphere \citep{pollack1965, muhleman1979, depater1991}. In this paper, we present microwave observations of Venus in the frequency bands centered near 40 and 90~GHz, corresponding to the altitudes of emission of approximately 35 and 50~km from the surface, respectively \citep{muhleman1979}. The measurements were performed using the Cosmology Large Angular Scale Surveyor (CLASS), a telescope array located at $22^\circ 58^\prime$~S latitude and $67^\circ 47^\prime$~W longitude in the Atacama Desert of northern Chile.

CLASS is a multifrequency polarimeter that surveys $\sim$~70\% of the microwave sky at large angular scales with the aim of measuring the primordial gravitational wave background and constraining the optical depth due to cosmic reionization \citep{tom2014,katie2016}. While CLASS is primarily designed to observe the cosmic microwave background (CMB) polarization, its high sensitivity allows it to observe other microwave sources within its survey area. In addition to its nominal CMB survey mode, CLASS periodically observes on-sky calibration sources including the Moon, Venus, and Jupiter to obtain telescope pointing information, characterize the beam response \citep{xu2020}, and calibrate the detector power response to the antenna temperature of the source \citep{appel19}. Between 25~August 2018 and 11~October 2018, the CLASS 40~GHz (\textit{Q} band) and 90~GHz (\textit{W} band) telescopes performed 74 dedicated Venus observations. The same instruments observed Jupiter 70 times between 26~June 2020 and 6~August 2020. In this paper, we use Jupiter as a calibration source to constrain the brightness temperature of Venus. We also examine the phase dependence of the measured Venus brightness temperature throughout the observing campaign. We describe the observations and results in Section \ref{sec:observations}, followed by the implications of the results in  Section \ref{sec:discussion}.

\section{Observations} \label{sec:observations}
During dedicated planet observations, we scan the telescope across the source over a small range ($\pm 14^\circ$) of azimuth angle at a fixed elevation of $45^\circ$ while the source rises or sets through the telescope field of view ($\sim 10^\circ$ in radius). As the telescope scans across the source, we obtain time-ordered data (TOD) for each detector at $\sim$~200~samples per second. During analysis, the raw TOD is first calibrated to measured optical power and then combined with the telescope pointing information to produce source-centered maps in telescope coordinates. Refer to \citet{appel19} and \citet{xu2020} for further details on CLASS data acquisition, calibration from raw detector data to observed optical power, and map-making from dedicated observations.

Given the telescope beam sizes (FWHM of $\sim 1.5^\circ$ for \textit{Q} band and $\sim 0.6^\circ$ for \textit{W} band), both Venus and Jupiter ($\lesssim$ 1$^\prime$ in angular diameter) are well approximated as point sources for both CLASS instruments. A point source's brightness temperature ($T_\mathrm{s}$) relates to the peak response as measured above the atmosphere by CLASS detectors ($T_\mathrm{m}^*$) as:
\begin{equation}
    T_\mathrm{s} \Omega_\mathrm{s} = T_\mathrm{m}^* \Omega_\mathrm{B},
\label{eq:scaling}
\end{equation}
where $\Omega_\mathrm{B}$ is the beam solid angle and $\Omega_\mathrm{s}$ is the solid angle subtended by the source \citep{page2003a}. To obtain $T_\mathrm{m}^*$ for \textit{W}-band detectors, we correct the actual measured response $T_\mathrm{m}$ for atmospheric transmission to account for the effect of precipitable water vapor (PWV) at the CLASS site. For each observation, we obtained PWV data from
the Atacama Pathfinder Experiment/the Atacama Cosmology Telescope (APEX/ACT)\footnote{For Venus observations, we use APEX PWV data obtained from \url{https://archive.eso.org/wdb/wdb/asm/meteo\_apex/form}. Since the APEX radiometer was offline during our Jupiter observations, we
instead use ACT PWV data (acquired via private correspondence) obtained from a 183 GHz radiometer \citep{bustos2014} operated by Universidad Cat\'{o}lica de la Sant\'{i}sima Concepci\'{o}n.} and used the Atacama Large Millimeter/submillimeter Array atmospheric transmission model\footnote{\url{https://almascience.eso.org/about-alma/atmosphere-model}} based on the ATM code described in \citet{padro2001} to calculate the transmission correction factor for each detector.
At \textit{W} band, this correction was necessary because $T_\mathrm{m}^*$ is greater than $T_\mathrm{m}$ by up to $\sim$ 5\% for observations with high ($\sim$ 5~mm) PWV. However, at \textit{Q} band, the effect of PWV on the derived brightness temperature ratio (Section \ref{sec:temp}) for a detector is $\lesssim$ 1\%, which is within our measurement uncertainty; therefore no PWV-related correction was applied, i.e., $T_\mathrm{m}^*$ = $T_\mathrm{m}$. 

To increase the signal-to-noise ratio of the measurement, we average the source-centered maps from individual observations to form an aggregate map of the source per detector. Since $\Omega_\mathrm{s}$ changes between observations, the averaging is done relative to a fiducial solid angle $\Omega_\mathrm{ref}$. This is achieved by scaling $T_\mathrm{m}^*$ from each observation by a factor of $\Omega_\mathrm{ref}/\Omega_\mathrm{s}$ while averaging the maps. To determine $\Omega_\mathrm{s}$ subtended by a planet, we use the distance to the planet that varies for each observation and a fixed disk radius ($R$). For Jupiter, we calculate an effective $R$ for the projected area of its oblate disk using the method described in \citet{weiland2011}. For Venus, we use a standard disk radius of 6120~km \citep{muhleman1979, fahd1992, butler2001}, which includes the physical surface radius of Venus ($\sim$ 6052~km) plus the height of the atmosphere. This choice of Venus disk radius allows our results to be compared with previous measurements and the brightness temperature models presented in Section~\ref{sec:discussion}.

\begin{figure*}
\begin{center}
\includegraphics[scale=0.465]{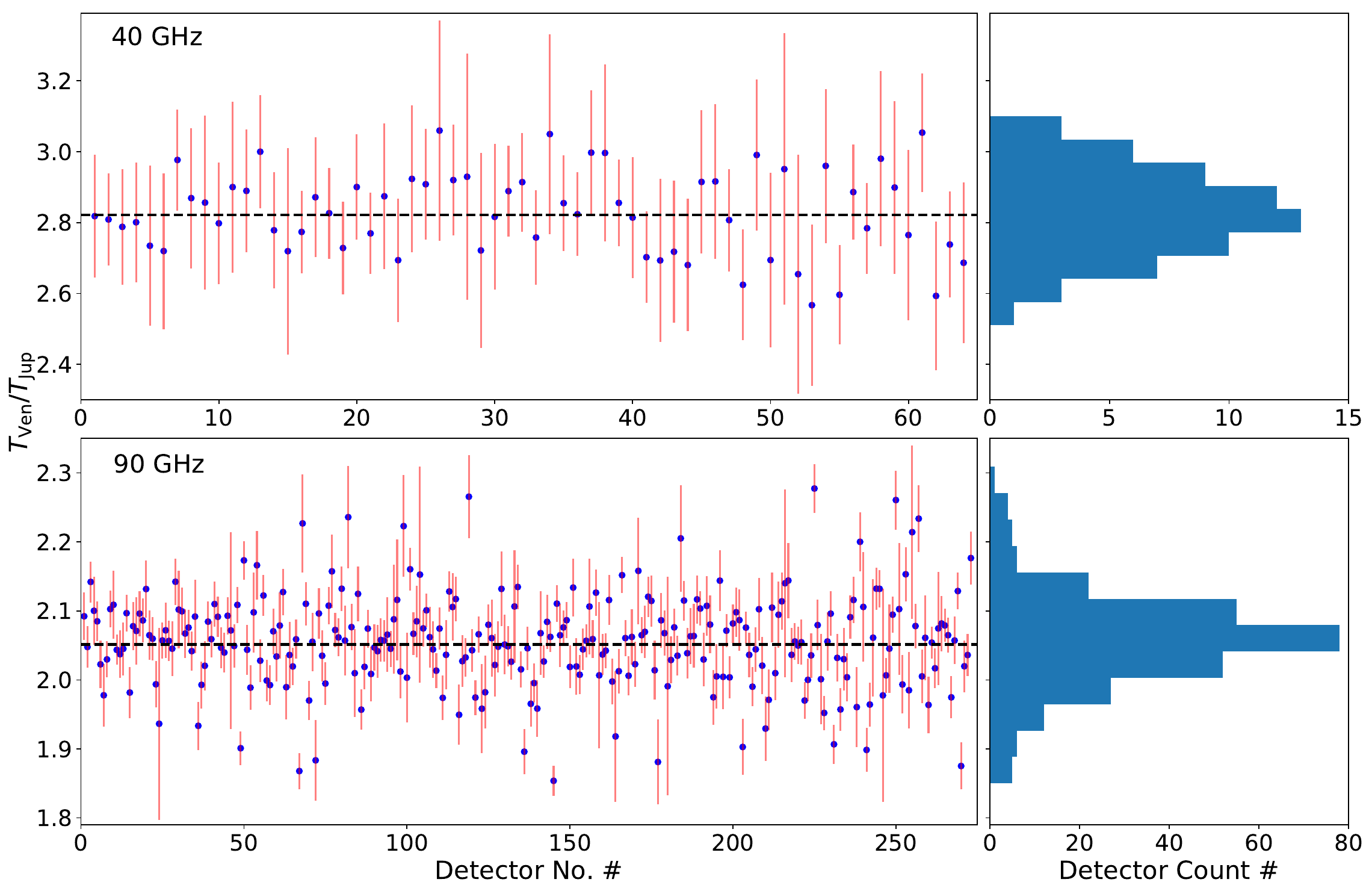}
\end{center}
\caption{(Left) Venus-to-Jupiter brightness temperature ratios measured by the CLASS 40 and 90 GHz detectors. Each data point corresponds to the result obtained from the averaged maps for a particular detector. For a given detector, the brightness temperature ratio was calculated by
taking the ratio of measured peak amplitudes of Venus and Jupiter, both scaled to the same fiducial reference solid angle of 5.5 $\times$~10$^{-8}$ sr. The errors in the ratios are the combined errors from the Venus and Jupiter amplitude measurements, which were calculated from the variance of baseline measurements away from the source. The inverse-variance weighted mean ratios (dashed lines) for the 40 and 90 GHz detectors are 2.821 $\pm$ 0.015 and  2.051 $\pm$ 0.004, respectively. (Right) Corresponding histograms of the brightness temperature ratio measurements.}
\label{fig:ven_temp}
\end{figure*}

\subsection{Brightness Temperature}\label{sec:temp}
For averaged planet maps, we can write Equation~\ref{eq:scaling} as $T_\mathrm{s} \Omega_\mathrm{ref} = T_\mathrm{ref} \Omega_\mathrm{B}$, where $T_\mathrm{ref}$ is the average of ($T_\mathrm{m}^*$ $\times~\Omega_\mathrm{ref}/\Omega_\mathrm{s}$) values acquired over the observing campaign for a particular detector. We set $\Omega_\mathrm{ref}$ to be the same for both Venus and Jupiter per-detector averaged maps so that the ratio of peak responses measured by CLASS detectors $T_\mathrm{ref}(\mathrm{Ven})/T_\mathrm{ref}(\mathrm{Jup})$ is equal to the ratio of brightness temperatures $T_\mathrm{Ven}/T_\mathrm{Jup}$. We calculated $T_\mathrm{ref}(\mathrm{Ven})$ and $T_\mathrm{ref}(\mathrm{Jup})$ values for $\Omega_\mathrm{ref}$ = 5.5 $\times$~10$^{-8}$~sr (i.e., 54.55$^{\prime\prime}$ diameter) to obtain the $T_\mathrm{Ven}/T_\mathrm{Jup}$ ratios shown in Figure \ref{fig:ven_temp}. For 40~GHz (90~GHz), out of 71 (331) detectors that detected both Venus and Jupiter, we consider 64 (273) of them with
$T_\mathrm{Ven}/T_\mathrm{Jup}$ between 2.5 and 3.2 (1.8 and 2.4) and uncertainty in the ratio $<$ 15\% (10\%) for this analysis. These particular choices of data filters were applied to reject the detectors that were affected by excess noise or improper calibration from raw TOD to optical power. The uncertainties in the ratios shown in Figure \ref{fig:ven_temp} are the combined errors from the Venus and Jupiter peak amplitude (in units of optical power) measurements, which were obtained from the variance of baseline measurements away from the source and include the calibration error from raw TOD to optical power. With about 5\% calibration error per measurement (J.~W. Appel et al. 2021, in preparation) and $\sim$~50 maps averaged per detector, the contribution from statistical calibration error is $\sim$~0.7\% of the measured amplitude. 

For the 40 and the 90 GHz detector arrays, the inverse-variance weighted mean $T_\mathrm{Ven}$/$T_\mathrm{Jup}$ ratios are \mbox{2.821 $\pm$ 0.015} and \mbox{2.051 $\pm$ 0.004}, respectively, where the uncertainties are the standard errors on the mean. Using bootstrapping, we verified that these standard errors represent the uncertainties in the mean of the underlying distributions. For both frequency bands, the standard deviation of the mean values of $10^6$ bootstrap-generated resamples (the statistic converges well before $10^6$ resamplings) is same as the standard error of the parent sample.  

To obtain Venus brightness temperatures, we multiply these CLASS-measured mean $T_\mathrm{Ven}$/$T_\mathrm{Jup}$ ratios by $T_\mathrm{Jup}$ values from the Wilkinson Microwave Anisotropy Probe (WMAP), calibrated with respect to the CMB dipole. \citet{bennett2013} reported nine-year mean $T_\mathrm{Jup}$ values of \mbox{$154.3 \pm 0.59$}~K and \mbox{172.8 $\pm$ 0.52}~K at \textit{Q} and \textit{W} bands centered at \mbox{40.78 $\pm$ 0.07}~GHz and \mbox{93.32 $\pm$ 0.19}~GHz, respectively. The effective Rayleigh-Jeans (RJ) point source center frequencies for CLASS \textit{Q} and \textit{W} band detectors are \mbox{38.8 $\pm$ 0.2}~GHz and \mbox{93.7 $\pm$ 0.2}~GHz, respectively (S. Dahal et al. 2021, in preparation). To correct for the difference in effective center frequencies between CLASS and WMAP detectors at \textit{Q} band, we utilize two different methods to compute $T_\mathrm{Jup}$ at 38.8~GHz: (1) a local power law fit between WMAP's Ka and \textit{Q} band measurements gives \mbox{$T_\mathrm{Jup} = 152.56 \pm 0.61$}~K, and (2) the nominal Radio BErkeley Atmospheric Radiative transfer (RadioBEAR)\footnote{\url{https://github.com/david-deboer/radiobear}} model provides \mbox{$T_\mathrm{Jup}$} of 152.58~K. We adopt our local power law fit to the WMAP $T_\mathrm{Jup}$ spectrum for further analysis, since although the RadioBEAR model agrees well with the WMAP measurements at \textit{Q} and V bands, the model is $\sim$~0.7\% higher at Ka band. It is notable that the close proximity of the CLASS and WMAP \textit{Q}-band center frequencies results in agreement between the two methods at 38.8~GHz. At \textit{W} band, since the CLASS and WMAP effective center frequencies are close with overlapping error bars, the frequency correction is not necessary. Combined with the CLASS-measured mean $T_\mathrm{Ven}$/$T_\mathrm{Jup}$ ratios, these $T_\mathrm{Jup}$ values yield the Venus brightness temperature of 430.4~$\pm$~2.8~K at \textit{Q} band and 354.5~$\pm$~1.3~K at \textit{W} band. 

These brightness temperature values are measured with respect to blank sky; they do not include the CMB contribution blocked by the planet, which is included in the background. Therefore, to get the absolute RJ brightness temperature of Venus, we add the RJ temperature of the CMB: 1.9~K at \textit{Q} band and 1.1~K at \textit{W} band \citep{fixsen1996, page2003a}. As a result, we get the absolute Venus brightness temperature of 432.3~$\pm$~2.8~K and 355.6~$\pm$~1.3~K at \textit{Q} and \textit{W} bands, respectively. To our knowledge, these results are the most precise disk-averaged Venus brightness temperature measurements to date at both frequency bands.

\subsection{Phase}\label{sec:phase}
\begin{figure}[ht]
\begin{center}
\includegraphics[scale=0.465]{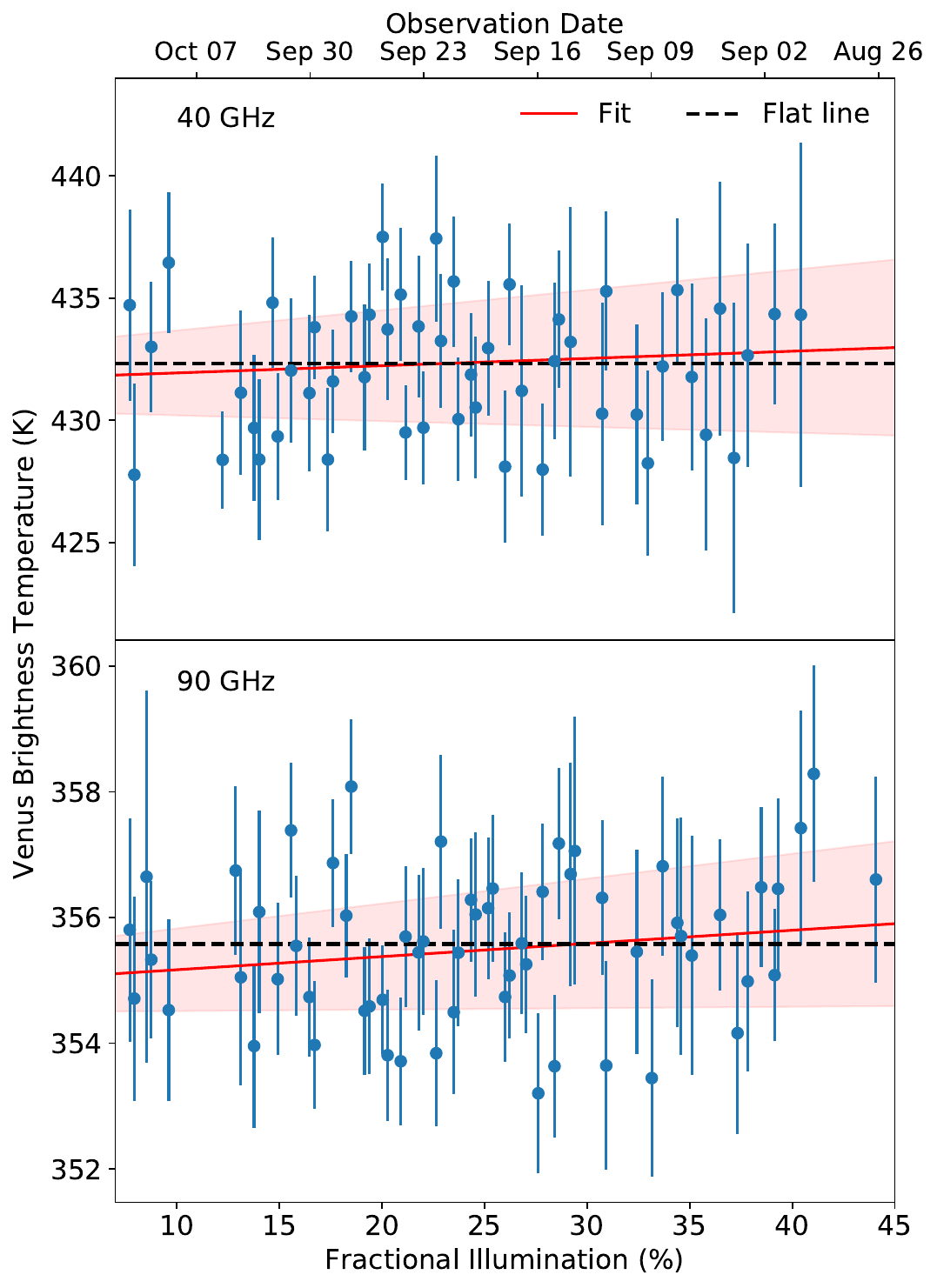}
\end{center}
\caption{Measured brightness temperature versus fractional solar illumination (phase) of Venus during the observing campaign. Each data point corresponds to the detector array-averaged temperature value for that particular day. While, over time, the fractional illumination decreases from 44\% to 8\%, we do not observe any statistically significant phase dependence of the measured temperatures. The best fit lines (solid red) correspond to a gradient of  $0.03 \pm 0.05$~K/\% and $0.02 \pm 0.02$~K/\%  for the 40 and the 90 GHz observations, respectively. The shaded regions show 1$\sigma$ uncertainties for the fits. For comparison, the flat dashed lines show the absolute brightness temperature values calculated in Section~\ref{sec:temp}.}
\label{fig:venus_phase}
\end{figure}

During the CLASS Venus observing campaign, the solar illumination of Venus changed from 44\% to 8\% (with 100\% illumination occurring at superior conjunction). To examine the phase dependence of the Venus brightness temperature, we calculate the detector array-averaged temperature values for individual observations (i.e., before averaging the maps) using the Jupiter-based calibration discussed in Section \ref{sec:temp}. As discussed earlier in this section, the individual \textit{W}-band measurements have been corrected for the effect of PWV. Figure~\ref{fig:venus_phase} shows the array-averaged Venus brightness temperature values plotted against the fractional solar illumination and its corresponding observation date. During this observing period, we detect no phase dependence of the Venus brightness temperature; the gradient of the array-averaged temperature values for different solar illuminations is statistically consistent with a flat line for both frequency bands. The best fit lines have gradients of $0.03 \pm 0.05$~K/\% and $0.02 \pm 0.02$~K/\%  for the 40 and the 90 GHz frequency bands, respectively.

\section{Discussion}\label{sec:discussion}
\begin{figure}[ht]
\begin{center}
\includegraphics[scale=0.37]{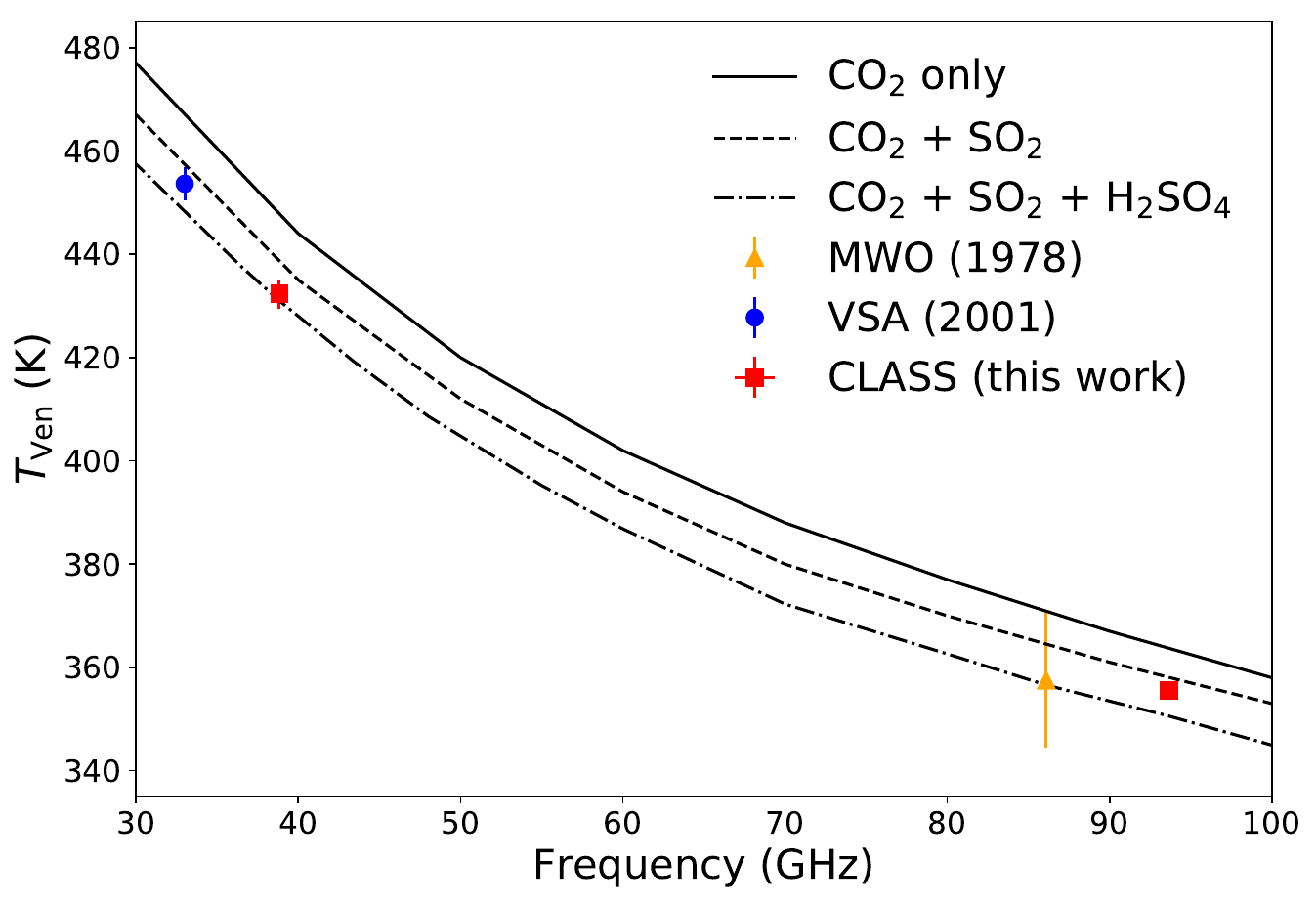}
\end{center}
\caption{Microwave spectrum of Venus. We compare the CLASS measurements with previous measurements from the Millimeter Wave Observatory (MWO; \citealt{ulich1980}) at \textit{W} band and the Very Small Array (VSA; \citealt{hafez2008}) at Ka band. We also plot three different brightness temperature models: CO$_2$ only (solid) and CO$_2$+SO$_2$ (dashed) from  \citet{fahd1992}, and CO$_2$+SO$_2$+H$_2$SO$_4$ (dashed-dotted) from \citet{akins2020}. In the CO$_2$+SO$_2$ model, the SO$_2$ abundance is set to 62~ppm below the cloud base and exponentially decreasing above it. For the model with H$_2$SO$_4$, the sub-cloud SO$_2$ abundance is set to 50 ppm whereas the gaseous H$_2$SO$_4$ abundance ranges from \mbox{10--14}~ppm following the equatorial profile presented in \citet{kolodner1998}. CLASS measurements are inconsistent with the CO$_2$-only model and prefer the presence of trace amounts of additional absorbers at both frequency bands.}
\label{fig:venus_comparison}
\end{figure}

The microwave thermal emission from Venus is strongly affected by its atmospheric opacity. Venusian atmospheric models \citep{depater1991, fahd1992} predict that the opacity provided by CO$_2$ alone leads to brightness temperatures of 444 K and 367 K at 40 and 90 GHz frequencies, respectively. Any presence of additional absorbers like SO$_2$ and H$_2$SO$_4$ increases the opacity of Venus at these frequencies. This moves the radiative transfer weighting function higher in the atmosphere to colder altitudes \citep{depater1991}, hence decreasing the predicted brightness temperature. In Figure \ref{fig:venus_comparison}, we compare our measurements to three different Venusian brightness temperature models: (1)  CO$_2$ only, (2) CO$_2$ with SO$_2$ models from \citet{fahd1992}, and (3) CO$_2$ with both SO$_2$ and H$_2$SO$_4$ (gaseous) from \citet{akins2020}. In the second model, the abundance profile for SO$_2$ is set to 62 ppm below 48~km altitude (the lower cloud base) and exponentially decreasing above 48~km. For the third model with both additional components, a uniform subcloud SO$_2$ abundance of 50~ppm along with an H$_2$SO$_4$ abundance ranging from 10--14 ppm following the equatorial profile from \citet{kolodner1998} is used. While the temperature and pressure profiles used in all three models are based on probe measurements over low Venusian latitudes, we note that the difference between the models with and without H$_2$SO$_4$ could be partly attributed to the difference in assumed profiles. Refer to \citet{fahd1992} and \citet{akins2020} for further details on these models.

At both frequency bands, our measurements agree with a CO$_2$-dominant atmospheric model. Furthermore, given their precision, our measured brightness temperature values favor the presence of trace amounts of additional absorbers. As shown in Figure \ref{fig:venus_comparison}, our measured brightness temperature is consistent with the H$_2$SO$_4$ model at \textit{Q} band and prefers a slightly lower H$_2$SO$_4$ abundance at higher altitudes probed by \textit{W}-band frequencies. Our data also supports a CO$_2$+SO$_2$ model with SO$_2$ concentration slightly higher than 62 ppm. However, identifying the exact content and abundance of these additional absorbers requires further modeling of the Venusian atmosphere and is outside the scope of this work. We also note that the CO$_2$-only model from \citet{fahd1992} is inconsistent ($> 5\sigma$) with our measurements at both frequencies. In Figure~\ref{fig:venus_comparison}, we also compare our values with two previous measurements in this frequency range. Our brightness temperature values are consistent with the \citet{ulich1980} measurement at 86.1 GHz and the \citet{hafez2008}\footnote{\citet{hafez2008} reported $T_\mathrm{Ven}$ = \mbox{$460.3 \pm 3.2$}~K relative to blank sky, $T_\mathrm{Jup}$ = 146.6 K, and Venus angular size based on its surface radius. We use an updated $T_\mathrm{Jup}$ = 147.1 K from \citet{bennett2013} and correct for the standard Venus disk radius and CMB contribution to obtain the final absolute \mbox{$T_\mathrm{Ven}$ = 453.6 $\pm$ 3.1}~K.} measurement at 33~GHz.

The lack of phase dependence in our measured brightness temperatures is also in agreement with the insignificant temperature variation throughout 1.5 synodic cycles of Venus reported by \citet{hafez2008}. Previously, \citet{basharinov1965} reported \mbox{$T_\mathrm{Ven} = 427 + 41 \cos (\Phi - 21^\circ)$~K}, where \mbox{$-180^\circ < \Phi < 180^\circ$} is the Venus phase angle, based on their 37.5 GHz observations near superior conjunction (\mbox{$\Phi = 0^\circ$}). To explain this observation, \citet{pollack1965} used an atmospheric model with dust distributed through the lower atmosphere with preferential abundance in the illuminated hemisphere, which could lead to the approximately 10\% variation in the \textit{Q}-band brightness temperature amplitude relative to the Venus phase. This is inconsistent with both the \citet{hafez2008} observations and the results presented in this paper. Our measured brightness temperature at \textit{Q} band, which agrees with the CO$_2$-dominant atmospheric model (Figure \ref{fig:venus_comparison}), also does not require the dust-filled model from \citet{pollack1965}. The lack of phase dependence at \textit{Q} band may suggest strong enough winds in the CO$_2$-dominant lower atmospheric layers to evenly distribute the temperature around the planet. This would also explain the lack of phase dependence we observe at \textit{W} band. However, pinpointing the exact mechanism behind the lack of phase dependence at these microwave frequencies requires further investigation.

Both CLASS \textit{Q}- and \textit{W}-band telescopes continue to observe the microwave sky from the Atacama Desert in Chile. Future data from these instruments will provide improved phase coverage of the Venusian illumination, which can help further characterize the phase-dependent temperature variation (if any) at these microwave frequencies. Furthermore, CLASS has started observations with a dichroic instrument  \citep{dahal2020} operating at frequency bands centered near 150 and 220~GHz. The addition of this high-frequency data will also improve our frequency coverage in the Venus spectrum, providing better understanding of the Venusian atmosphere. 

\section{Summary}\label{sec:summary}
We present Venus observations performed with the \textit{Q}-band and the \textit{W}-band CLASS telescopes. Using Jupiter as a calibration source, we measure the disk-averaged absolute brightness temperature of Venus to be \mbox{$432.3 \pm 2.8$}~K and \mbox{$355.6 \pm 1.3$}~K in the \textit{Q} and \textit{W} bands centered at \mbox{$38.8 \pm 0.2$}~GHz and \mbox{$93.7 \pm 0.2$}~GHz, respectively. These results are the most precise Venus brightness temperature measurements to date. At both frequency bands, the measured brightness temperature values are consistent with CO$_2$-dominant atmospheric models \citep{depater1991, fahd1992, akins2020} and previous measurements from \citet{ulich1980} and  \citet{hafez2008}. During our two-month observing campaign, while the fractional solar illumination of Venus changed from 44\% to 8\%, we did not observe any phase dependence of the brightness temperature in either frequency band. This lack of phase dependence agrees with the measurement from \citet{hafez2008} and does not support the brightness temperature versus phase relation at \textit{Q} band reported by \citet{basharinov1965}.

\section*{Acknowledgments}
We acknowledge the National Science Foundation Division of Astronomical Sciences for their support of CLASS under Grant Numbers 0959349, 1429236, 1636634, 1654494, and 2034400. We thank Johns Hopkins University President R. Daniels and the Deans of the Kreiger School of Arts and Sciences for their steadfast support of CLASS. We further acknowledge the very generous support of Jim and Heather Murren (JHU A\&S '88), Matthew Polk (JHU A\&S Physics BS~'71), David Nicholson, and Michael Bloomberg (JHU Engineering~'64). The CLASS project employs detector technology developed in collaboration between JHU and Goddard Space Flight Center under several previous and ongoing NASA grants. Detector development work at JHU was funded by NASA grant number NNX14AB76A. We acknowledge scientific and engineering contributions from Max Abitbol, Fletcher Boone, David Carcamo, Ted Grunberg, Saianeesh Haridas, Connor Henley, Lindsay Lowry, Nick Mehrle, Isu Ravi, Daniel Swartz, Bingjie Wang, Qinan Wang, Tiffany Wei, and Zi\'ang Yan. We thank William Deysher, Miguel Angel D\'iaz, and Chantal Boisvert for logistical support. We acknowledge productive collaboration with Dean Carpenter and the JHU Physical Sciences Machine Shop team.

S.D. is supported by an appointment to the NASA Postdoctoral Program at the NASA Goddard Space Flight Center, administered by the Universities Space Research Association under contract with NASA.
S.D. acknowledges support under NASA-JHU Cooperative Agreement 80NSSC19M005. R.B. acknowledges support for the 183 GHz radiometer from the UCSC project DINREG 06/2017. R.R. acknowledges partial support from CATA, BASAL grant AFB-170002, and CONICYT-FONDECYT through grant 1181620. Z.X. is supported by the Gordon and Betty Moore Foundation. CLASS is located in the Parque Astron\'omico Atacama in northern Chile under the auspices of the Agencia Nacional de Investigaci\'on y Desarrollo (ANID). We thank Imke de Pater, Darrell Strobel, and Alex~B. Akins for discussions that improved this work.

\software{\texttt{PyEphem} \citep{rhodes2011}, \texttt{NumPy} \citep{numpy}, \texttt{SciPy} \citep{scipy}, \texttt{Astropy} \citep{astropy}, \texttt{Matplotlib} \citep{matplotlib}}, \texttt{RadioBEAR} \citep{radiobear1,radiobear2}

\bibliography{venus}{}
\bibliographystyle{aasjournal}
\end{document}